\documentclass[preprint,12pt]{elsarticle}

\usepackage{graphics,epsfig}
\usepackage{graphicx}
\usepackage{float}
\usepackage{amssymb,amstext,amsmath}
\usepackage{mathtools}
\usepackage{booktabs}
\usepackage{natbib}
\usepackage[latin1]{inputenc}
\usepackage{epstopdf}

\DeclareMathOperator{\tr}{tr}

\begin{document}

\begin{frontmatter}
\title{Formation of grains and dislocation structure of geometrically necessary boundaries}
\author{M. Koster, K. C. Le\footnote{corresponding author: +49 234 32-26033, email: chau.le@rub.de.}}
\address{Lehrstuhl f\"{u}r Mechanik - Materialtheorie, Ruhr-Universit\"{a}t Bochum,\\D-44780 Bochum, Germany}
\begin{abstract} 
A continuum dislocation model of formation of grains whose boundaries have a non-vanishing thickness is proposed. For a single crystal deforming in simple shear the lamellar structure of grains with thin layers containing dislocations as the geometrically necessary boundaries turns out to be energetically preferable. The thickness and the energy of this type of grain boundary are computed as functions of the misorientation angle.
\end{abstract}

\end{frontmatter}

\section{Introduction}
\label{sec:Introduction}
One of the main guiding principles in seeking an appropriate theory of formation of grains in metals and alloys during and after cold working processes producing severe plastic deformations has first been proposed by Hansen and Kuhlmann-Wilsdorf \cite{Hansen1986} in form of the so-called LEDS-hypothesis: the dislocation structures in the final state of deformation minimize the energy of crystals (see also \cite{Kuhlmann1989,Kuhlmann1991,Laird1986}). The main reason why the formation of grains becomes energetically preferable at severe plastic deformations lies in the non-convexity of the energy of crystal in this range \cite{Carstensen2002,Koster2015,Ortiz99,Ortiz00}. Within the conventional crystal plasticity considered in \cite{Carstensen2002,Ortiz99,Ortiz00} the minimization of such non-convex energy leads immediately to the infinitely fine lamellar structure with grain boundaries as sharp interfaces. However, as mentioned by Kuhlmann-Wilsdorf and Hansen \cite{Kuhlmann1991}, typical grain boundaries, termed  geometrically necessary boundaries, have as a rule a non-vanishing thickness and may contain a large number of dislocations and thus contradict the conventional crystal plasticity. The question then arrises in this connection: what kind of continuum model may resolve this conflict? The present  paper proposes a dislocation model of formation of grains within the continuum dislocation theory \cite{Berdichevsky2006a,Le2014nonlinear} which predicts the existence of such geometrically necessary boundaries. By including the energy of dislocation network containing the gradient of the plastic slip into the energy functional we regularize the non-convex energy minimization problem. We illustrate the application of the theory on the example of single crystal deforming in single slip under a simple shear. We show that the geometrically necessary boundaries, in which the transition from one grain to the next occurs smoothly, have a small but finite thickness and contain a large number of dislocations. Although the resultant Burgers vector of dislocations in such grain boundary is non-zero, they do not produce long range stresses, and the lamellar structure of grains is in fact the low energy dislocations structure. We also compute the thickness of geometrically necessary boundaries and their energies as functions of the misorientation angles. Based on these results we estimate also the number of grains in terms of the specimen sizes.

\section{Continuum theory of formation of grains}
We consider for simplicity an initially dislocation-free single crystal having only one active slip system. In this case the kinematic quantities characterizing its observable deformations are the placement field $\mathbf{y}(\mathbf{x})$ and the plastic slip field $\beta (\mathbf{x})$. The incompatible plastic deformation is given by
\begin{equation*}
\mathbf{F}^p (\mathbf{x})=\mathbf{I}+ \beta (\mathbf{x}) \mathbf{s}\otimes \mathbf{m},
\end{equation*}  
with the pair of constant and mutually orthogonal unit lattice vectors $\mathbf{s}$ and $\mathbf{m}$ denoting the slip direction and the normal to the slip planes. Using the multiplicative resolution of the total compatible deformation gradient $\mathbf{F}=\partial \mathbf{y}/\partial \mathbf{x}$ into the plastic and elastic part \cite{Le2014nonlinear}, we find the incompatible elastic deformation in the form
\begin{equation*}
\mathbf{F}^e=\mathbf{F}\cdot \mathbf{F}^{p-1}=\frac{\partial \mathbf{y}}{\partial \mathbf{x}}\cdot (\mathbf{I}- \beta \mathbf{s}\otimes \mathbf{m}).
\end{equation*} 
The tensor of dislocation density measuring the incompatibility of $\mathbf{F}^p$ reads (see \cite{Le2014nonlinear,Ortiz99})
\begin{equation*}
\mathbf{T}=-\mathbf{F}^p\times \nabla =\mathbf{s}\otimes (\nabla \beta \times \mathbf{m}).
\end{equation*}
If, in addition, all dislocation lines are straight lines parallel to the unit vector $\mathbf{l}$, then the scalar dislocation density (or the number of excess dislocations per unit area perpendicular to $\mathbf{l}$) can be determined as
\begin{equation*}
\rho =\frac{|\mathbf{T\cdot \mathbf{l}|}}{b}=\frac{1}{b}|(\nabla \beta \times \mathbf{m})\cdot \mathbf{l}|,
\end{equation*}
with $b$ being the magnitude of Burgers vector.

For crystals having as a rule small elastic strains we propose the free energy per unit volume of the undeformed configuration in the form
\begin{equation}\label{eq:1.4}
\psi (\mathbf{E}^e,\rho ) = \frac{1}{2}\lambda (\tr \mathbf{E}^e)^2 + \mu \tr(\mathbf{E}^e\cdot \mathbf{E}^e) +\frac{1}{2} \mu k\frac{\rho ^2}{\rho _s^2}.
\end{equation}
Here $\mathbf{E}^e=\frac{1}{2}(\mathbf{F}^{eT}\cdot \mathbf{F}^e-\mathbf{I})$ corresponds to the elastic strain tensor, $\lambda $ and $\mu $ are the Lam\'{e} constants, $k$ a material constant, and $\rho _s$ can be interpreted as the saturated dislocation density. The first two terms in \eqref{eq:1.4} represent energy of crystal due to the macroscopic elastic deformation. The last term describes energy of the dislocation network for moderate dislocation densities. Note that for small or extremely large dislocation densities close to the saturated density the logarithmic energy proposed in \cite{Berdichevsky2006b} is more appropriate. We deform this crystal occupying in the initial configuration some region $\mathcal{V}$ of three-dimensional space by placing it in a displacement-controlled device such that, at the boundary $\partial \mathcal{V}$, the conditions
\begin{equation}\label{eq:1.6}
\mathbf{y}(\mathbf{x})=\bar{\mathbf{F}}\cdot \mathbf{x},\quad \beta (\mathbf{x})=0 \quad \text{for $\mathbf{x}\in \partial \mathcal{V}$}
\end{equation}
are specified, with $\bar{\mathbf{F}}$ being a given overall deformation. If the deformation process is isothermal, no body force acts on this crystal, and the resistance to the dislocation motion can be neglected, then the following variational principle turns out to be valid: the true placement vector $\check{\mathbf{y}}(\mathbf{x})$ and the true plastic slip $\check{\beta }(\mathbf{x})$ in the {\it final} equilibrium state of deformation minimize the energy functional
\begin{equation}
\label{eq:1.7}
I[\mathbf{y}(\mathbf{x}),\beta (\mathbf{x})]=\int_{\mathcal{V}}w(\mathbf{F},\beta ,\nabla \beta )\, dx
\end{equation}
among all {\it continuously differentiable} fields $\mathbf{y}(\mathbf{x})$ and $\beta (\mathbf{x})$ satisfying constraints \eqref{eq:1.6}, where $w(\mathbf{F},\beta ,\nabla \beta )=\psi (\mathbf{E}^e,\rho )$. We will see that, due to the non-convexity of the free energy density \eqref{eq:1.4} and the presence of $\nabla \beta $ in the energy functional via the energy of the dislocation network, the formation of grains with regular grain boundaries having a finite thickness is energetically preferable.

\section{Energy minimizer in plane strain simple shear}
\label{sec:SimpleShear}
\begin{figure}[htb]
    \begin{center}
    \includegraphics[width=9cm]{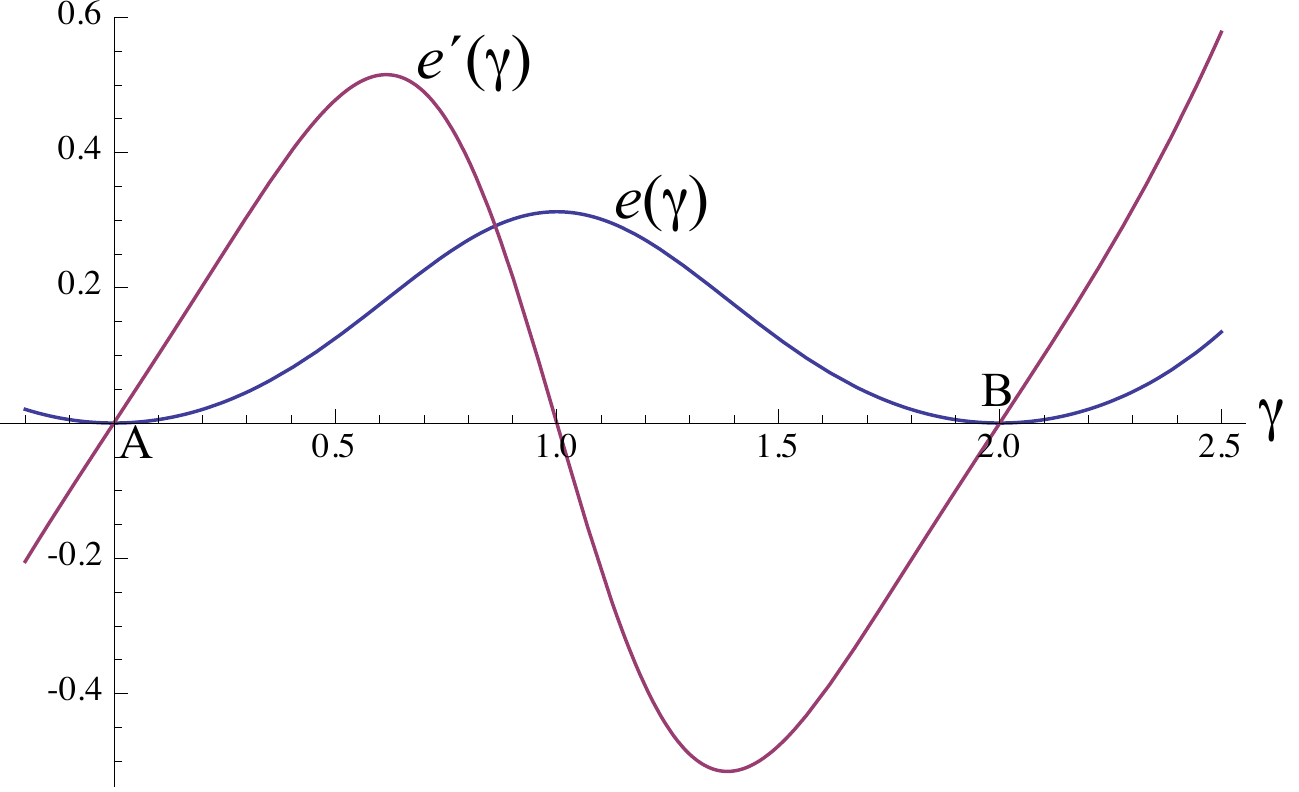}
    \end{center}
    \caption{Condensed energy $e(\gamma )$ and the dimensionless shear stress $e^\prime (\gamma )$ for $\varphi =-45^\circ$.}
    \label{fig:condensed}
\end{figure}

Consider now the special case of plane strain simple shear of the specimen in form of a cuboid of height $H$, width $L$, and depth $D$ such that $y_3=x_3$, while $y_1(\mathbf{x})$, $y_2(\mathbf{x})$ and $\beta (\mathbf{x})$ depend only on two cartesian coordinates $x_1$ and $x_2$ and satisfy at the side boundary the conditions
\begin{equation*}
y_1=x_1+\gamma x_2, \quad y_2=x_2,\quad \beta =0, 
\end{equation*}
with $\gamma$ being the overall shear strain. We assume that $\mathbf{s}^T=(\cos \varphi ,\sin \varphi ,0)$, $\mathbf{m}^T=(-\sin \varphi ,\cos \varphi ,0)$ and all dislocation lines are parallel to the $x_3$-axis, so $\rho =|\nabla \beta \cdot \mathbf{s}|/b$. If the deformations are uniform such that
\begin{equation*}
\mathbf{F} =\bar{\mathbf{F}}= \mathbf{I}+\gamma \mathbf{e}_1\otimes \mathbf{e}_2, 
 \quad \mathbf{F}^p = \mathbf{I}+\beta \mathbf{s}\otimes \mathbf{m},
\end{equation*}
with  $\gamma $ and $\beta $ being the constants, then the energy \eqref{eq:1.7} normalized by $\mu |\mathcal{V}|$ and minimized with respect to $\beta $ turns out to be non-convex for $\varphi \in (-\pi /2,0)$ as shown in Fig.~\ref{fig:condensed} (see \cite{Koster2015}). 

In view of this non-convexity, we proposed in \cite{Koster2015} the energy minimizing sequence consisting of layers having the uniform states A and B according to
\begin{align*}
&\gamma _A=0, &  \gamma _B=-2\cot \varphi ,
\\
&\beta _A=0, &  \beta _B=2\cot \varphi ,
\end{align*}
such that the volume fraction of the layer $B$ is given by $s=-\gamma /(2\cot \varphi )$. It has been shown in \cite{Koster2015} that such candidates for the minimizer satisfy the equations of equilibrium in each layer as well as the outer boundary conditions except at the side boundaries $x_1=0,L$ of the specimen. Besides, the energy of such lamellar structure is equal to zero which is the minimal possible value. However, if the boundaries between layers are sharp interfaces, these candidates do not belong to the set of admissible fields of our original variational problem \eqref{eq:1.6} and \eqref{eq:1.7} due to the jumps of $\mathbf{F}$ and $\beta $ on those interfaces, so they fail to be the energy minimizers of \eqref{eq:1.6} and \eqref{eq:1.7}.

\begin{figure}[htb]
    \begin{center}
    \includegraphics[width=12 cm]{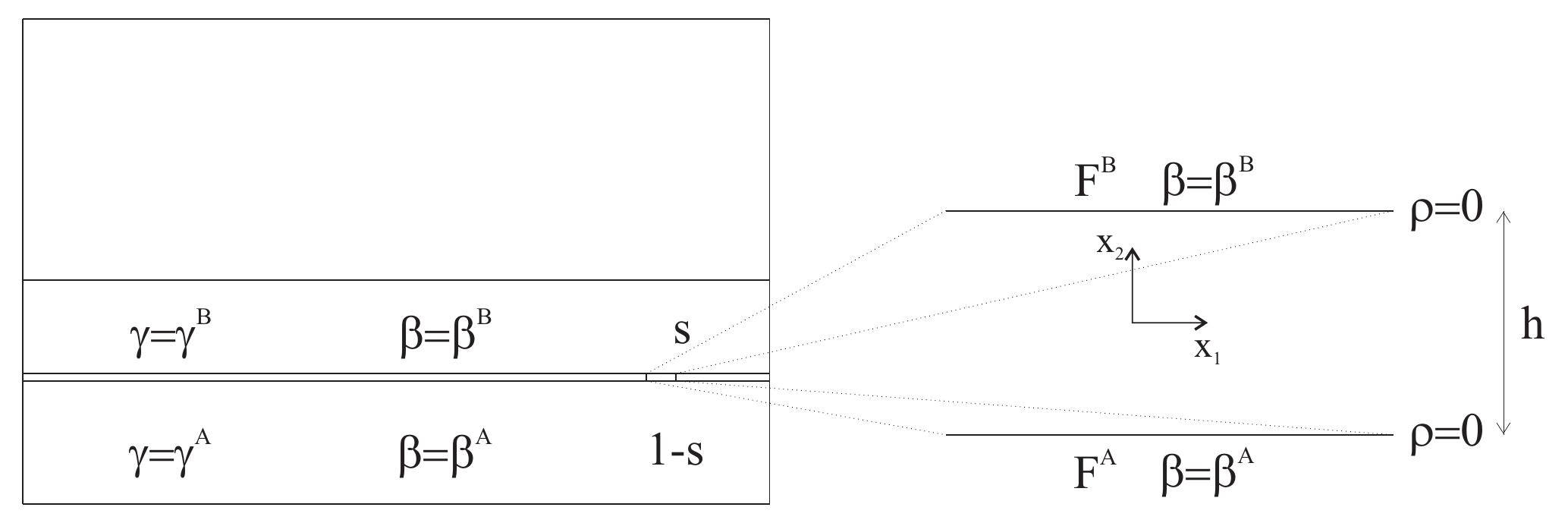}
    \end{center}
    \caption{Enhanced grain boundary model.}
    \label{fig:grainBoundaryLayer}
\end{figure}

To correct the behavior of those candidates for minimizers we assume now that the layers corresponding to the states A and B are separated by a thin layer of small thickness $h$ in which the placement and plastic slip change smoothly from state A to state B (see Fig.~\ref{fig:grainBoundaryLayer}). Since this boundary layer is thin ($h$ is much smaller than $sH$, $(1-s)H$, and the sizes of the specimen), it is reasonable to assume that the displacement in $x_2$-direction is zero, while the displacement in $x_1$-direction and the plastic slip depend only on $x_2$: 
$$y_1=x_1+u(x_2),\quad y_2=x_2,\quad \beta =\beta (x_2).$$ 
With this Ansatz it is easy to show that the determination of functions $u(x_2)$ and $\beta (x_2)$ as well as the unknown boundary layer reduces to minimizing the following functional 
\begin{equation}\label{eq:2.3}
I_b \left[ u(x_2),\beta (x_2)\right] =  \int_{l_1}^{l_2}\left[f\left(u_{,2},\beta\right) + g\left(\beta_{,2}\right)\right] dx_2
\end{equation}
among functions $u(x_2)$, $\beta (x_2)$ and unknown lengths $l_1$, $l_2$ such that
\begin{align}
u_{,2}(l_1)&=\gamma _A=0, &  u_{,2}(l_2)&=\gamma _B=-2\cot \varphi , \notag
\\
\beta (l_1)&=\beta _A=0, &  \beta (l_2)&=\beta _B=2\cot \varphi . \label{eq:2.3a}
\end{align}
Although the analytical solution can be found in the most general case with $\lambda \ne 0$, we choose $\lambda =0$ for the shortness of presentation. In this special case $f\left(u_{,2},\beta\right)$ and $g(\beta _{,2})$ are given by
\begin{align*}
&f\left(u_{,2},\beta\right) =  \frac{1}{4} \left\{\left[\beta^2 \sin^4 \varphi + \left(\beta u_{,2} \sin^2 \varphi + \beta \sin\varphi \cos\varphi + 1\right)^2 - 1\right]^2\right. \notag
\\
&\left. + \left[\left(\beta u_{,2} \sin\varphi \cos\varphi + \beta \cos^2 \varphi - u_{,2}\right)^2 + \left(\beta \sin\varphi \cos\varphi - 1\right)^2 - 1\right]^2\right. 
\\
&\left. + 2 \left[\left(-\beta u_{,2} \sin\varphi \cos\varphi - \beta \cos^2 \varphi + u_{,2}\right) \left(\beta u_{,2} \sin^2\varphi + \beta \sin\varphi \cos\varphi + 1\right) \right.\right. \notag
\\
&\left.\left. + \beta \sin^2 \varphi \left(1 - \beta \sin\varphi \cos\varphi\right)\right]^2\right\}, \quad
g\left(\beta_{,2}\right) = \frac{1}{2} k \frac{\beta_{,2}^2\sin ^2\varphi }{b^2\rho_s ^2}. \notag
\end{align*}
Functional \eqref{eq:2.3} can be reduced to the functional depending only on $\beta $. Indeed, the variation of \eqref{eq:2.3} with respect to $u(x_2)$ leads to the equation and boundary conditions that imply
\begin{equation*}
\frac{\partial f}{\partial u_{,2}}=0, 
\end{equation*}
so the stresses inside the boundary layer are zero. Since it was already established in \cite{Koster2015} that the stresses in layers A and B are also zero, the whole specimen is stress-free. Solving the above equation with respect to $u_{,2}$ we express it in terms of $\beta $
\begin{equation*}
u_{,2}\left(\beta\right)=\frac{\beta\left(\beta \sin 2\varphi - 2\cos 2\varphi\right)}{\beta^2\left(\cos 2\varphi - 1\right) + 2\beta \sin 2\varphi - 2}.
\end{equation*}
Substituting this back into the functional \eqref{eq:2.3} we reduce the latter to the functional depending only on $\beta (x_2)$ and $l_1$, $l_2$
\begin{equation}\label{eq:2.6}
I_b\left[ \beta (x_2)\right] =  \int_{l_1}^{l_2}\left[p\left(\beta\right) + g\left(\beta_{,2}\right)\right] dx_2,
\end{equation}
where
\begin{align*}
p(\beta )&=f(u_{,2}(\beta),\beta )= \beta^2 \sin^2 \varphi  \left(\beta \sin\varphi - 2\cos\varphi\right)^2 \\
&\left\{ \frac{3 \beta^4 - 8\beta^3 \sin 2\varphi + 4\beta^3 \sin 4\varphi - 4 \left(\beta^2 + 2\right) \beta^2 \cos 2\varphi}{8\left[\beta^2 \left(1 - \cos 2\varphi\right) + \beta^2 - 2 \beta \sin 2\varphi + 2\right]^2}\right.\\
&\left. + \frac{\left(\beta^2 - 4\right) \beta^2 \cos 4\varphi + 12\beta^2 - 16\beta \sin 2\varphi + 16}{8\left[\beta^2 \left(1 - \cos 2\varphi\right) + \beta^2 - 2 \beta \sin 2\varphi + 2\right]^2}\right\} .
\end{align*}
Varying functional \eqref{eq:2.6} with respect to $\beta (x_2)$, $l_1$, $l_2$ and taking into account the constraints \eqref{eq:2.3a} we obtain the Euler equation
\begin{equation}
\label{eq:2.7}
 \frac{d}{dx_2}g^\prime (\beta _{,2})-p^\prime (\beta )=0
\end{equation}
which is subjected to the boundary conditions
\begin{equation}
\label{eq:2.8}
\begin{split}
\beta (l_1)=0, \quad  \beta (l_2)=2\cot \varphi ,
\\
\beta _{,2}(l_1)=0, \quad \beta _{,2}(l_2)=0.
\end{split}
\end{equation}
\begin{figure}[htb]
    \begin{center}
		\includegraphics[width=9cm]{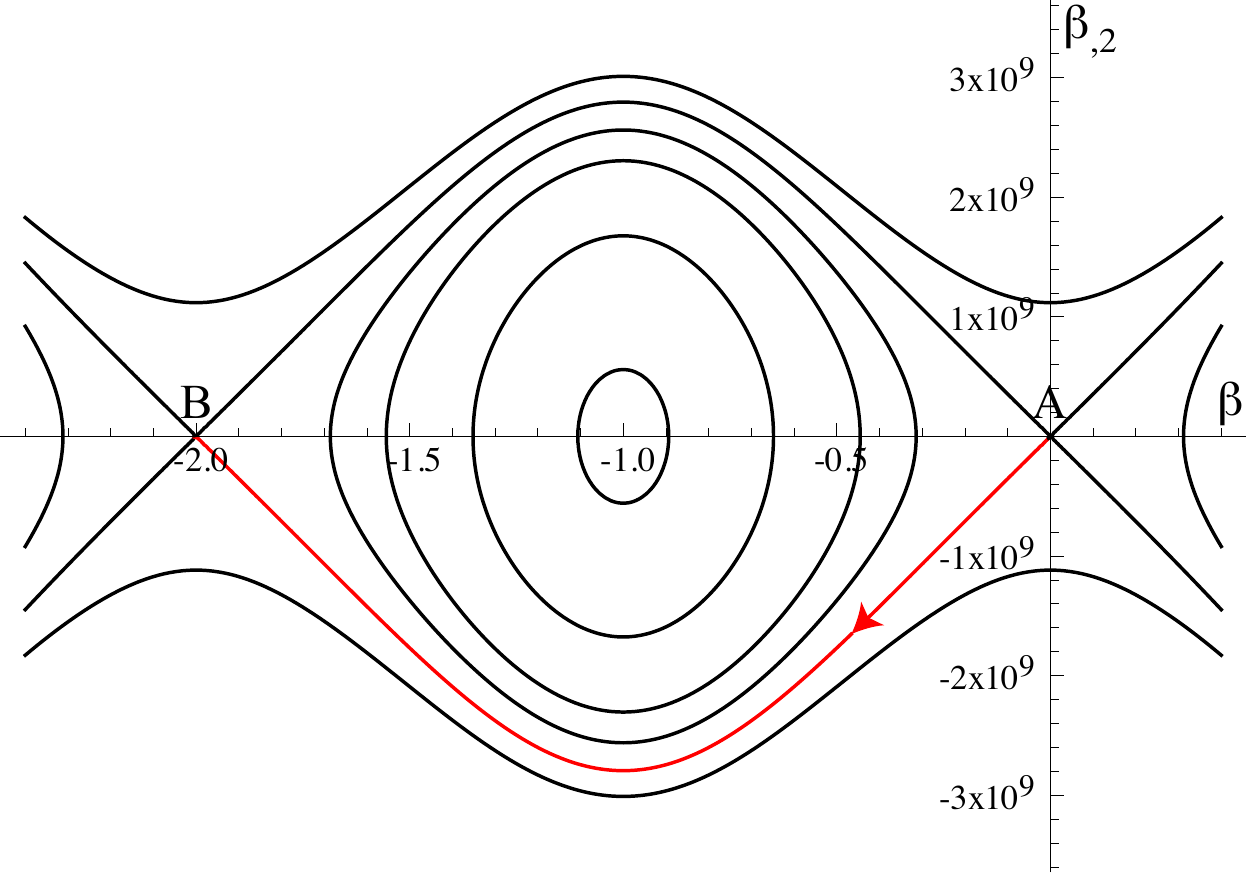}
    \end{center}
    \caption{Phase portrait and the separatrix connecting the states A and B.}
    \label{fig:phasePortrait}
\end{figure}
The last two boundary conditions mean that the dislocation density must vanish at the boundaries $x_2=l_1$ and $x_2=l_2$. Equation \eqref{eq:2.7} possesses the first integral
\begin{equation}\label{eq:2.11}
g(\beta _{,2})-p(\beta ) = c.
\end{equation}
Using the boundary conditions \eqref{eq:2.8} as well as the identities $p(\beta _A)=p(\beta _B)=0$, we find that $c=0$. The phase portrait computed for $b=2.5\times 10^{-10}$m, $\rho _s=10^{16}/$m$^2$, $k=10^{-6}$ and $\varphi =-\pi /4$ is shown in Fig.~\ref{fig:phasePortrait}. We see that the phase curve connecting the states A and B and satisfying the boundary conditions \eqref{eq:2.8} is the separatrix in the lower half of the phase plane $(\beta ,\beta _{,2})$ represented by the red curve with an arrow denoting the direction of change of $\beta $ as $x_2$ changes from $l_1$ to $l_2$. It is also easy to show that the phase curve connecting B and A is the separatrix in the upper half of the phase plane, so the dislocations in the boundary layer between B and A have another sign than those connecting A and B.

\begin{figure}[htb]
    \begin{center}
		\includegraphics[width=9cm]{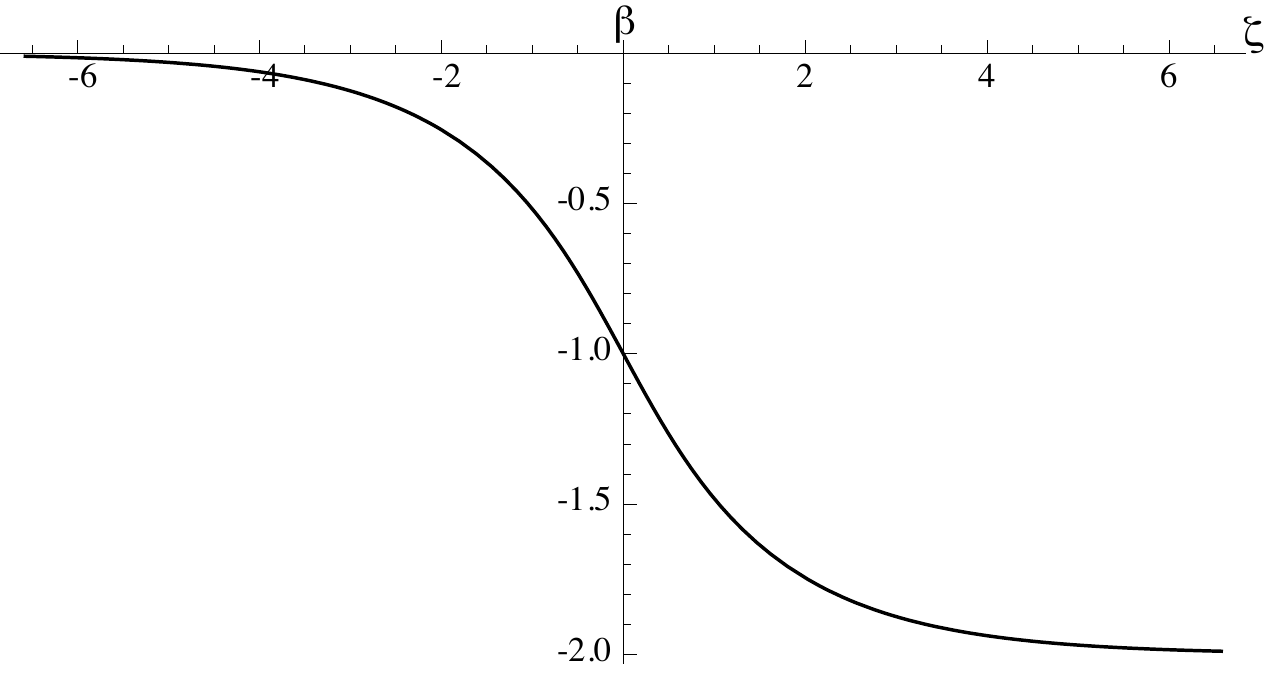}
    \end{center}
    \caption{Plot of $\beta $ versus the dimensionless coordinate $\zeta =\frac{\sqrt{2}x_2b\rho _s}{\sqrt{k}|\sin \varphi |}$.}
    \label{fig:beta}
\end{figure}

The first integral \eqref{eq:2.11} with $c=0$ enables one to find $\beta $ implicitly in terms of $x_2$ according to
\begin{equation*}
x_2=-\frac{\sqrt{k}|\sin \varphi |}{\sqrt{2}b\rho _s}\int_{\cot \varphi }^{\beta }\frac{dt}{\sqrt{p(t)}}.
\end{equation*}
The plot of $\beta $ versus the dimensionless coordinate $\zeta =\sqrt{2}x_2b\rho _s/(\sqrt{k}|\sin \varphi |)$ for the above chosen parameters is shown in Fig.~\ref{fig:beta}. It is seen that, for large $|\zeta |$, the plastic slip remains in the very close neighborhood of the state A with $\beta =0$ or the state B with $\beta =2\cot \varphi $, while the strong change of $\beta $ leading to the transition from A to B occurs in a finite interval of $\zeta $.

\begin{figure}[htb]
    \begin{center}
		\includegraphics[width=8cm]{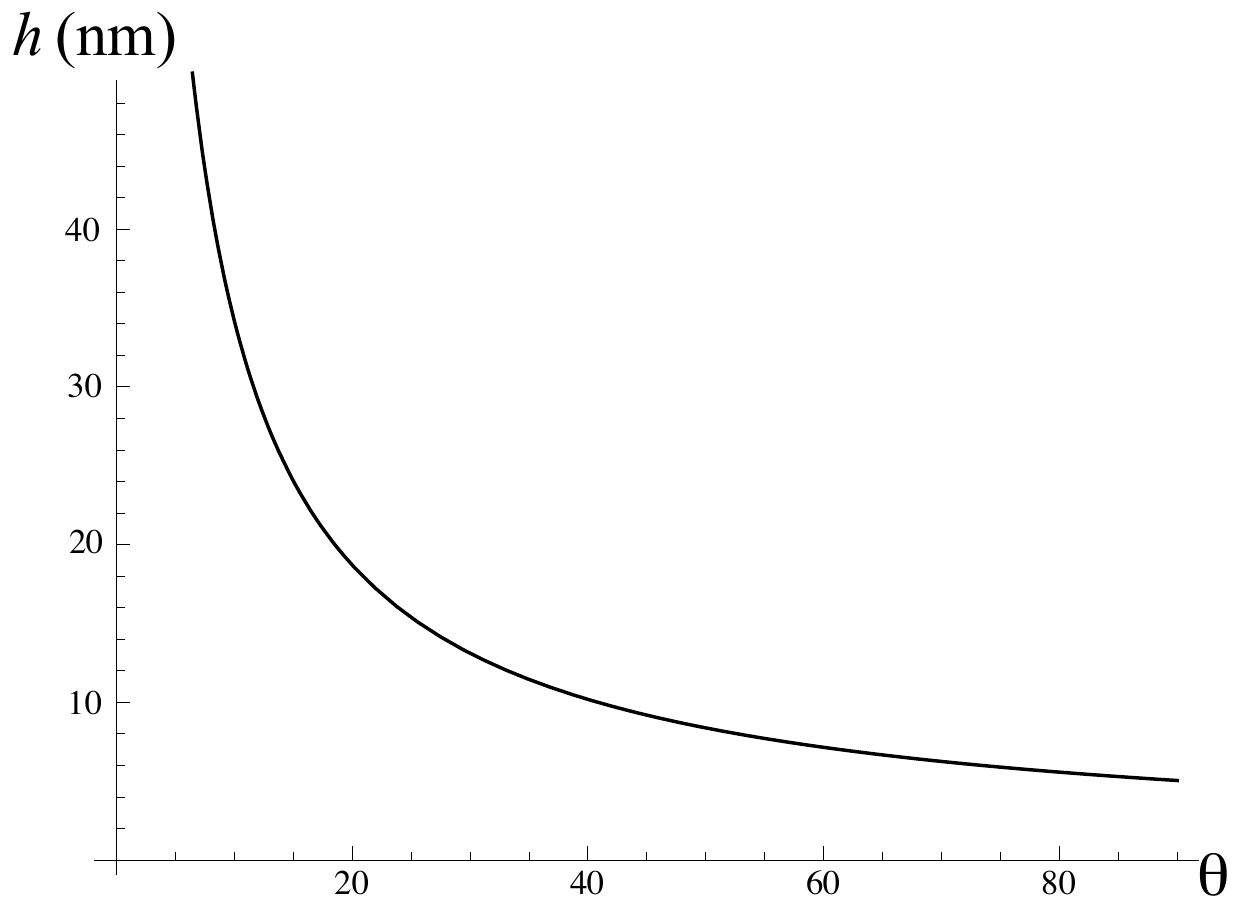}
    \end{center}
    \caption{Thickness of geometrically necessary boundary $h$ (in nanometer) versus the misorientation angle $\theta =2\varphi +\pi $ (in degrees).}
    \label{fig:thickness}
\end{figure}

It turns out that if we compute the thickness of the boundary layer according to the formula
\begin{equation*}
h=l_2-l_1=\frac{\sqrt{k}|\sin \varphi |}{\sqrt{2}b\rho _s}\int_{2\cot \varphi }^{0}\frac{dt}{\sqrt{p(t)}},
\end{equation*}
then this thickness becomes infinite, so the result contradicts our assumption about the smallness of $h$ as compared with $sH$, $(1-s)H$, and with the sizes of the specimen. The resolution of this conflict should be found in the discreteness of crystals that may accommodate only a large but finite natural number of dislocations. Since the dislocation density is $\rho =|\beta _{,2} \sin \varphi |/b$, the total number of dislocations in the boundary layer equals
\begin{equation*}
N=L\int_{l_1}^{l_2}\rho dx_2=|\beta (l_2)||\sin \varphi |\frac{L}{b}.
\end{equation*}
As the smallest number of dislocation is 1, the smallest quantum of plastic slip leading to the recognizable change of $N$ must be $\beta _q=b/(L|\sin \varphi |)$. Now, if we take $b/L$ as a small positive number and compute the thickness of the boundary layer according to
\begin{equation}\label{eq:2.13}
h=l_2-l_1=\frac{\sqrt{k}|\sin \varphi |}{\sqrt{2}b\rho _s}\int_{2\cot \varphi +\beta _q}^{-\beta _q}\frac{dt}{\sqrt{p(t)}},
\end{equation}
then the thickness becomes finite (except for $\varphi =0$). Since the pre-factor, the integrand, and both limits of integration depend on the orientation of slip system $\varphi $, the thickness is also a function of $\varphi $. Fig.~\ref{fig:thickness} shows the dependence of the thickness of geometrically necessary boundary (measured in nanometer) on the misorientation angle $\theta =2\varphi +\pi $ (measured in degrees, see \cite{Koster2015}) for the above chosen parameters and $b/L=10^{-4}$ (the result turns out to be not strongly sensitive to the choice of $b/L$). It is seen that the thickness changes from 50 to 6 nanometers for  misorientation angles larger than 10$^\circ$. For misorientation angles close to zero formula \eqref{eq:2.13} is not well-defined. In this case the dislocation model of small-angle tilt boundary considered in \cite{Read1950} is more appropriate.

\begin{figure}[htb]
    \begin{center}
		\includegraphics[width=7cm]{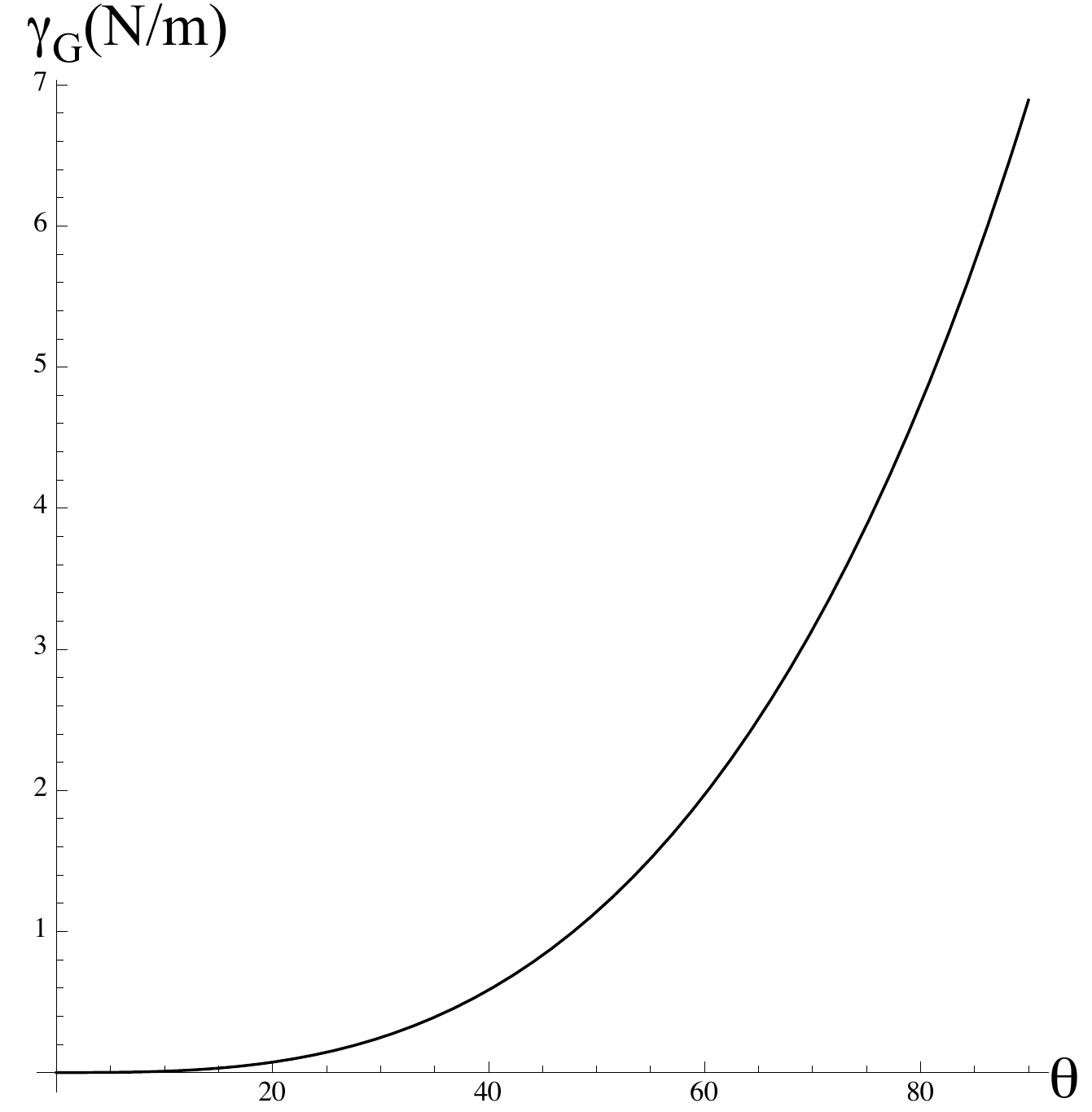}
    \end{center}
    \caption{Energy density of of geometrically necessary boundary $\gamma _G$ (in N/m) versus the misorientation angle $\theta =2\varphi +\pi $ (in degrees).}
    \label{fig:GBenergy}
\end{figure}

The minimum of the functional \eqref{eq:2.6} can easily be computed based on the analytical solution found above. Using the first integral we obtain
\begin{equation}\label{eq:2.14}
I_b = 2\int_{l_1}^{l_2}p(\beta )dx_2=\frac{2\sqrt{k}|\sin \varphi |}{\sqrt{2}b\rho _s}\int_{2\cot \varphi }^{0} \sqrt{p(\beta )}d\beta .
\end{equation}
Since this minimum value has the meaning of the energy (normalized by $\mu $) per unit area of the geometrically necessary boundary, the energy density of such boundary must be $\gamma _G =\mu I_b$. Similar to \eqref{eq:2.13}, the pre-factor, the integrand, and the lower limit of integration  in \eqref{eq:2.14} depend on the orientation of slip system $\varphi $, so $\gamma _G$ must also be a function of $\varphi $. Fig.~\ref{fig:GBenergy} shows the dependence of $\gamma _G$ (measured in N/m) on the misorientation angle $\theta =2\varphi +\pi $ (measured in degrees) for the above chosen parameters and $\mu =26$GPa (for aluminum). In contrast to $h$, the energy density of geometrically necessary boundary is well-defined for all misorientation angles. For small misorientation angles up to 50$^\circ$ the energy density of the geometrically necessary boundary changes from zero to approximately 1 N/m that agrees quite well with the value 0.625 N/m given in \cite{Hirth1968}. 

In \cite{Koster2015} we gave the estimation of the number of grains based on the following deliberations. The energy of the boundary layers near the side boundaries required to satisfy the side boundary conditions turns out to be of the order $\mu HD\varepsilon $, where $\varepsilon $ is the thickness of one pair of layers A and B. Taking into account the energy of the geometrically necessary boundaries which is of the order $\gamma _G DLH/\varepsilon $, we can estimate $\varepsilon $ by minimizing these two contributions to the energy yielding 
$$\varepsilon \sim \sqrt{\gamma _G L/\mu} \sim \sqrt{I_bL}.$$
This relation exhibits clearly the size effect.

\section{Conclusion}
\label{sec:Conclusion}
We have shown in this paper that the presence of the gradient of plastic slip in the energy of the dislocation network enables one to regularize the non-convex energy minimization in the class of smooth displacements and plastic slips. This leads to the formation of grains with geometrically necessary boundaries having a finite thickness. Let us mention the similarity of the proposed continuum theory with the static version of Cahn-Hilliard theory of phase separation \cite{Cahn1958} and the recent phase-field approach \cite{Chen2002}. The generalization of our model to capture the motion of geometrically necessary boundaries and the grain growth will be addressed elsewhere.

\bigskip
\noindent {\it Acknowledgments}

The financial support by the German Science Foundation (DFG) through the research project GP01-G within the Collaborative Research Center 692 (SFB692) is gratefully acknowledged.

\end{document}